\documentclass[10pt,journal,compsoc]{IEEEtran}
\usepackage{color}
\usepackage{graphicx}
\usepackage{url}
\usepackage{subfigure}

\usepackage{amsmath,amssymb,amsfonts}
\usepackage{algorithmic}
\usepackage{graphicx}
\usepackage{textcomp}
\usepackage{amssymb}
\usepackage{multirow}
\usepackage{array}
\usepackage{stfloats}
\usepackage[flushleft]{threeparttable}
\usepackage{booktabs}
\usepackage{bm}
\usepackage{booktabs}
\usepackage{hyperref}
\usepackage{amssymb}
\usepackage{array}
\usepackage{multicol}
\usepackage{multirow}
\usepackage{stfloats}
\usepackage{threeparttable}
\usepackage[table,xcdraw]{xcolor}
\definecolor{ForestGreen}{RGB}{0,0,0}

\ifCLASSOPTIONcompsoc
  \usepackage[nocompress]{cite}
\else
  \usepackage{cite}
\fi
\ifCLASSINFOpdf
\else
\fi
\begin{document}
%
\title{ImageCAS: A Large-Scale Dataset and Benchmark for Coronary Artery Segmentation based on Computed Tomography Angiography Images}
%
%
%
%

\author{An Zeng, Chunbiao Wu, Wen Xie, Jin Hong, Meiping Huang, Jian Zhuang, Shanshan Bi, Dan Pan, Najeeb Ullah, Kaleem
Nawaz Khan, Tianchen Wang, Yiyu Shi, Xiaomeng Li, Guisen Lin, Xiaowei Xu

\thanks{This work was supported by the Science and Technology Planning Project of Guangdong Province, China (No. 2019B020230003), Guangdong Peak Project (No. DFJH201802), the National Natural Science Foundation of China (No. 62006050, No. 62276071), Science and Technology Projects in Guangzhou, China (No. 202206010049, No. 2019A050510041), Guangdong Basic and Applied Basic Research Foundation (No. 2022A1515010157, 2022A1515011650), Guangzhou Science and Technology Planning Project (No. 202102080188), the Sanming Project of Medicine in Shenzhen (No. SZSM202011005), and Health Project, Guangdong  High-level Hospital Construction Fund.
}
\thanks{An Zeng and Biao Chun are with the School of Computer Science, Guangdong University of Technology, Guangzhou, China. Wen Xie, Jin Hong, Meiping Huang, Jian Zhuang and Xiaowei Xu are with Guangdong Provincial Key Laboratory of South China Structural Heart Disease, Guangdong Provincial People’s Hospital (Guangdong Academy of Medical Sciences), Southern Medical University, Guangzhou, China, 510080. (e-mail: xiao.wei.xu@foxmail.com)}
\thanks{Shanshan Bi is with the Department of Computer Science and Engineering, Missouri University of Science and Technology, Rolla, Missouri, United States. Pan Dan is with the Department of Computer Science, Guangdong Polytechnic Normal University, Guangzhou, China. Najeeb Ullah and Kaleem Nawaz Khan are with the Department of Computer Science, University of Engineering and Technology Mardan, KP, Pakistan. Tianchen Wang and Yiyu Shi is with the Department of Computer Science and Engineering, University of Notre Dame, Indiana, United States. Xiaomeng Li is with the Department of Electronic and Computer Engineering, The Hong Kong University of Science and Technology, Hong Kong SAR, China. Guisen Lin is with the Department of Radiology, Shenzhen Children's Hospital, Shenzhen, China.}
\thanks{An Zeng, Chunbiao Wu and Wen Xie contributed equally.}
\thanks{Xiaomeng Li, Guisen Lin and Xiaowei Xu are corresponding authors.}
}

%
%

\markboth{Journal of \LaTeX\ Class Files,~Vol.~14, No.~8, August~2018}%
{Shell \MakeLowercase{\textit{et al.}}: Bare Demo of IEEEtran.cls for Computer Society Journals}
%



\IEEEtitleabstractindextext{%
\begin{abstract}
Cardiovascular disease (CVD) accounts for about half of non-communicable diseases.
Vessel stenosis in the coronary artery is considered to be the major risk of CVD. 
Computed tomography angiography (CTA) is one of the widely used noninvasive imaging modalities in coronary artery diagnosis due to its superior image resolution. 
Clinically, segmentation of coronary arteries is essential for the diagnosis and quantification of coronary artery disease. 
Recently, a variety of works have been proposed to address this problem. 
However, on one hand, most works rely on in-house datasets, and only a few works published their datasets to the public which only contain tens of images. 
On the other hand, their source code have not been published, and most follow-up works have not made comparison with existing works, which makes it difficult to judge the effectiveness of the methods and hinders the further exploration of this challenging yet critical problem in the community. 
In this paper, we propose a large-scale dataset for coronary artery segmentation on CTA images. 
In addition, we have implemented a benchmark in which we have tried our best to implement several typical existing methods.
Furthermore, we propose a strong baseline method which combines multi-scale patch fusion and two-stage processing to extract the details of vessels.
Comprehensive experiments show that the proposed method achieves better performance than existing works on the proposed large-scale dataset. 
The benchmark and the dataset are published at https://github.com/XiaoweiXu/ImageCAS-A-Large-Scale-Dataset-and-Benchmark-for-Coronary-Artery-Segmentation-based-on-CT.

\end{abstract}

}

\maketitle

\IEEEdisplaynontitleabstractindextext

%
\IEEEpeerreviewmaketitle

\IEEEraisesectionheading{\section{Introduction}\label{sec:introduction}}

\IEEEPARstart{C}ardiovascular disease (CVD) is one of the leading health problems around the world nowadays. 
According to World Health Organization (WHO), 17.9 million deaths due to CVD occurred in 2019, accounting for 32\% of all global deaths \cite{world2009cardiovascular}.
Australian Institute of Health and Welfare (AIHW) reported that CVD was the leading cause of deaths in Australia, representing 42\% of all death in 2018 \cite{zhang2010coronary}. 
Among all the CVDs, coronary heart disease is the most common type  \cite{cooper2000trends} where the pathophysiology is mainly attributed to abnormal coronary artery stenosis.
Such stenosis often results in decreased myocardial perfusion and hypoxia damage of myocardial cells, and finally leads to myocardial infarction.
 
In clinical practice, computed tomography angiography (CTA) is widely used for diagnosis and treatment planning of coronary artery diseases since it is noninvasive and able to provides high-resolution 3D imaging \cite{collet2018coronary}. 
Once CTA images are obtained, radiologists first locate the coronary arteries and isolate its boundaries.
Then, the narrowing part is extracted and quantified for final diagnosis and treatment planning.
It is widely acknowledged that, however, such a process based on manual operations are unfavorably time-consuming 
and error-prone.
What is worse, the ever-increasing quantity and variety of medical images \cite{li2018large} can even make manual segmentation completely impracticable in terms of cost and reproducibility.
Therefore, automatic coronary artery segmentation is highly desirable.

However, this task is very challenging due to multiple reasons. 
First, the anatomical structure of coronary artery varies significantly from population to population.
For example, coronary arteries are usually surrounded by a layer of fat but in some people the arteries are within the heart muscle itself.
Second, some CTA images are noisy with artifacts that will cause low quality in segmentation result.
Third, the tubular structure of coronary arteries is extremely complex. For example, there is a large number of bifurcations along the arteries and a small proportion of coronary arteries area in the transverse planes \cite{zhu2021segmentation}. 

To address the above issues, tens of works have been proposed in the community in the last decade.
The approaches used in these works can be divided into two main categories: traditional machine learning (ML) based method, and deep learning (DL) based method.
Traditional ML-based method can be further sub-classified into pixel-based method \cite{doyle2006boosting,nguyen2012structure,tabesh2007multifeature,sirinukunwattana2015novel} and structure-based method \cite{altunbay2010color,gunduz2010automatic,fu2014novel,sirinukunwattana2015stochastic,lesage2016adaptive}.
These methods achieve promising results using hand-crafted features and prior knowledge of structures of coronary artery  \cite{zheng2011machine,mohr2012accurate,broersen2012frenchcoast,shahzad2013automatic,chi2015composite}.
However, they suffer considerable degradation when applied to coronaries with serious deformation.
Different from ML-based method, the recently proposed DL-based methods require little hand-crafted features or prior knowledge.
Works using such an approach has achieved considerable improvement over the first approach \cite{huang2018coronary,shen2019coronary,chen2019coronary,wolterink2019graph,kong2020learning,gu2021fusing,zhu2021segmentation,tian2021automatic}, proving its high effectiveness for coronary artery segmentation.

We perform a detailed analysis of existing works as shown in Table \ref{tab:works}, and find that most works fail to make a fair and comprehensive comparison with others.
For example, \cite{kong2020learning} did not make comparisons with any previous related works but just some popular deep neural networks (DNNs).
\cite{shen2019coronary} just listed the Dice score of related works for comparison which is not fair enough as the two methods were not evaluated using the same dataset.
Some works \cite{chi2015composite,han2016automatic} even used different sets of evaluation metrics tailored for specific clinical needs, while others \cite{shen2019coronary} even used different annotations in which the initial aorta is included, leading to a much higher Dice score. 
These evaluative biases are mainly due to the lack of a large-scale benchmark dataset available in the public domain.
The only two public datasets for coronary artery segmentation \cite{schaap2009standardized,kiricsli2013standardized} (also shown in Table \ref{tab:works}) only contain 8 and 18 images for training, respectively.
Besides that, all the existing methods have not released their source code, which has introduced more difficulties for a fair comparison.

In this paper, we propose a large dataset to fairly investigate the effectiveness of automatic coronary artery segmentation methods.
This dataset contains 1000 3D CTA images, which is considerably larger than the existing public datasets.
In addition, we also propose a benchmark based on this dataset, in which we not only implement several typical existing methods but also propose a strong baseline method.
Experimental results show that our baseline method achieves better performance than all the existing methods.
and meanwhile shows a good potential for improvement.
The contributions of this work are summarised as follows:
\begin{itemize}
\item We have collected a large-scale publicly available dataset containing 1,000 patients for coronary artery segmentation available at https://github.com/XiaoweiXu/ImageCAS-A-Large-Scale-Dataset-and-Benchmark-for-Coronary-Artery-Segmentation-based-on-CT.
\textcolor{ForestGreen}{An official data split of the dataset is also provided.}
We hope this dataset can help to promote related research in the community;
\item We have proposed a benchmark in which we implemented a variety of existing methods for coronary artery segmentation. We have also published the benchmark, and hope this could help follow-up works to make fair comparisons;
\item We have proposed a strong baseline method which combines multi-scale patch fusion and two-stage processing to extract the details of vessels, and the experimental results show that our method has outperformed existing state-of-the-art methods.

\end{itemize}
The rest of the paper is organized as follows. Section II gives an overview of related works.
In Section III, details of the proposed dataset is presented. 
Subsequently, Section IV presents the proposed benchmark including several typical existing methods and our baseline method.
The experimental results are presented and discussed in Section V, and Section VII concludes the paper.

 \begin{table*}[ht]
\caption{Existing traditional machine learning based works and their Dice scores (in \%) of coronary artery segmentation in the past decade.
In $a(b)$, $a$ is the overall quantity and $b$ is the quantity for training.
\textcolor{ForestGreen}{\textit{General} indicates that the method is designed for a variety of applications, while \textit{specific} indicates that the method is designed and optimized for coronary vessel segmentation.
}
}
\centering
\begin{tabular}{p{1cm}<{\centering}p{1cm}<{\centering}p{1cm}<{\centering}p{1cm}<{\centering}p{2cm}<{\centering}p{2cm}<{\centering}p{2cm}<{\centering}p{2cm}<{\centering}p{1cm}<{\centering}p{1cm}<{\centering}}\toprule[2pt]
\multirow{3}{*}{Work} & \multirow{3}{*}{Year}& \multirow{3}{*}{\begin{tabular}[c]{@{}c@{}}Data\\ quantity\end{tabular} }
&\multicolumn{2}{c}{\multirow{2}{*}{\begin{tabular}[c]{@{}c@{}}{Compared Traditional} \\{ML methods}\end{tabular}}}  
& \multicolumn{2}{c}{\multirow{2}{*}{\begin{tabular}[c]{@{}c@{}}{Compared} \\{Deep learning methods}\end{tabular}}}  &     
\multirow{3}{*}{\begin{tabular}[c]{@{}c@{}}Average\\ Dice score\end{tabular}} & 
\multirow{3}{*}{\begin{tabular}[c]{@{}c@{}}Code\\ available\end{tabular}} & 
\multirow{3}{*}{\begin{tabular}[c]{@{}c@{}}Data\\ available\end{tabular}}  
\\  
\\ 
&&& \multicolumn{1}{c}{General}  & \multicolumn{1}{c}{Specific}  & General & Specific \\ 

\midrule[2pt] 
\cite{zheng2011machine} & 2011 & 54(40)   &\cite{Kroon2022}&NA&NA&NA& NA & No & No 
\\\hline 
\cite{mohr2012accurate} & 2012 & 24$^\P$ &NA&NA&NA&NA& 70-73 & No & \cite{kiricsli2013standardized}  
\\\hline 
\cite{wang2012vessel} &2012 & 42(18)&NA&NA&NA&NA& 68-72 & No & \cite{kiricsli2013standardized}  
\\\hline 
\cite{broersen2012frenchcoast} & 2012 & 42(18) &NA&NA&NA&NA& 66 & No & \cite{kiricsli2013standardized}  
\\\hline 
\cite{shahzad2013automatic} & 2013 & 48(18) &NA&NA&NA&NA& 65 & No & \cite{kiricsli2013standardized}
\\\hline  %
\cite{lugauer2014improving}& 2014 &48(18)&NA&
\cite{mohr2012accurate}\cite{wang2012vessel}\cite{shahzad2013automatic} &NA&NA&72-74 &No &\cite{kiricsli2013standardized}
\\\hline 
\cite{lugauer2014precise}&2014 &48(18)&NA&\cite{mohr2012accurate} \cite{lugauer2014improving} \cite{shahzad2013automatic} \cite{wang2012vessel}&NA&NA&75-77&No&\cite{kiricsli2013standardized}
\\\hline 
\cite{chi2015composite} & 2015 & 10(6) &NA&NA&NA&NA& 84 & No & No     
\\ \hline 
\cite{lesage2016adaptive} & 2016 & 61(10) &\cite{skare2003improved} \cite{douc2009optimality}&NA&NA&NA& 86.2 & No & No 
\\\hline 
\cite{han2016automatic} & 2016 & 32(8) &NA&\cite{schaap2009rotterdam}&NA&NA& $<$84.3 & No & \cite{schaap2009standardized} 
\\\hline 

\cite{freiman2017improving} & 2017 &48(18)&NA&
\begin{tabular}[have l]{@{}l@{}}\cite{mohr2012accurate}\\ \cite{lugauer2014improving} \end{tabular}&NA&NA&69-74&No&\cite{kiricsli2013standardized}
\\\hline 
\cite{gao2019automatic} &2017 & 50$^\P$&NA&NA&NA&NA&93-95*&No&No
\\\hline 
\cite{du2021automated}& 2021 & 100$^\P$ &NA & NA &
\cite{han2014fast} \cite{yu2017automatic} \cite{shen2019coronary} & NA & 82 & No & No 
\\\bottomrule[2pt] 

\hline

\multicolumn{10}{l}{$^\P$ Only testing data is involved (training data is not applicable). }\\
\multicolumn{10}{l}{$*$ The segmentation result includes the initial part of the aorta. }\\

\end{tabular}


\label{tab:works}
\end{table*}

 \begin{table*}[ht]
\caption{Existing deep learning based works and their Dice scores (in \%) of coronary artery segmentation in the past decade. In $a(b)$, $a$ is the overall quantity and $b$ is the quantity for training.
\textcolor{ForestGreen}{\textit{General} indicates that the method is designed for a variety of applications, while \textit{specific} indicates that the method is designed and optimized for coronary vessel segmentation.
}}
\centering
\begin{tabular}{p{1cm}<{\centering}p{1cm}<{\centering}p{1cm}<{\centering}p{1cm}<{\centering}p{2cm}<{\centering}p{2cm}<{\centering}p{2cm}<{\centering}p{2cm}<{\centering}p{1cm}<{\centering}p{1cm}<{\centering}}\toprule[2pt]
\multirow{3}{*}{Work} & \multirow{3}{*}{Year}& \multirow{3}{*}{\begin{tabular}[c]{@{}c@{}}Data\\ quantity\end{tabular} }
&\multicolumn{2}{c}{\multirow{2}{*}{\begin{tabular}[c]{@{}c@{}}{Compared Traditional} \\{ML methods}\end{tabular}}}  
& \multicolumn{2}{c}{\multirow{2}{*}{\begin{tabular}[c]{@{}c@{}}{Compared} \\{Deep learning methods}\end{tabular}}}  &     
\multirow{3}{*}{\begin{tabular}[c]{@{}c@{}}Average\\ Dice score\end{tabular}} & 
\multirow{3}{*}{\begin{tabular}[c]{@{}c@{}}Code\\ available\end{tabular}} & 
\multirow{3}{*}{\begin{tabular}[c]{@{}c@{}}Data\\ available\end{tabular}}  
\\  
\\ 
&&& \multicolumn{1}{c}{General$^\P$}  & \multicolumn{1}{c}{Specific$^\S$}  & General & Specific \\ 

\midrule[2pt] 
\cite{moeskops2016deep} & 2016 & 10(6)  & NA & NA & NA & NA & $<$65$^{\ddag}$ &No & No  \\\hline

\cite{kjerland2017segmentation}& 2017 &42(18)& NA&NA&NA&NA&78&No& \cite{kiricsli2013standardized} \\\hline

\cite{duan2018context}&2018&50(40)&\cite{frangi1998multiscale} &NA &NA &NA & 80 &No&No\\\hline

\cite{chen2018automatic} & 2018 & 44(35) & NA & NA & NA & NA & 86 & No & No \\\hline 

\cite{huang2018coronary} & 2018 & 52(45)&NA&\cite{mohr2012accurate}\cite{shahzad2013automatic}&NA&\cite{moeskops2016deep}\cite{kjerland2017segmentation} & 83$^{\dag}$ & No & \cite{kiricsli2013standardized}     \\\hline

\cite{shen2019coronary} & 2019 & 70(50)& NA&NA&\cite{milletari2016v}&NA& 90* & No & No               \\

\cite{chen2019coronary} & 2019 & 15(11) &NA&NA&NA&NA& 80 & No & No                 \\\hline

\cite{mirunalini2019segmentation} & 2019 & 50(40) & NA & NA & NA & NA & 92 & No & No  \\\hline

\cite{wolterink2019graph} & 2019 & 42(18) &NA&\cite{lugauer2014precise}\cite{freiman2017improving}&NA&NA &73-75 & No & \cite{kiricsli2013standardized}                  \\\hline 

\cite{lee2019template} &2019 & 548(274) & NA & NA & \cite{ronneberger2015u}\cite{long2015fully}&NA&85&No&No\\\hline

\cite{yang2019discriminative} & 2019 & 8(7) &\cite{cetin2015higher}  \cite{aylward2002initialization}  \cite{friman2010multiple}
&NA&NA&NA&$<$99$\dag$&No&\cite{schaap2009standardized} \\\hline

\cite{fu2020mask}&2020&25(20)&NA&NA&NA&NA&90&No&No\\\hline

\cite{lei2020automated}&2020&48(18)&NA&NA&NA&NA&83&No&\cite{kiricsli2013standardized}\\\hline

\cite{gu2020segmentation} & 2020 & 70(50) & NA & NA & \cite{cciccek20163d}\cite{milletari2016v}\cite{zhang2019attention}\cite{long2015fully}\cite{huang20193d} & NA & 91 & No & No  \\\hline

\cite{kong2020learning} & 2020 & 916(733) &NA&NA&\cite{yu2017automatic}&NA& 85 & No & No                 \\  \hline

\cite{wang2021geometric}& 2021 & 48(18) & NA& NA& NA& \cite{moeskops2016deep}\cite{kjerland2017segmentation}\cite{chen2019coronary}&91&No&\cite{kiricsli2013standardized}\\\hline 

\cite{gu2021fusing} & 2021 & 59(34) &NA&NA&\cite{patravali20172d}\cite{duan2019automatic}\cite{mortazi2017cardiacnet}\cite{chen2018voxresnet}\cite{milletari2016v}&NA& 87 & No & No                \\  \hline
\cite{zhu2021segmentation} & 2021 & 30(24) &NA&NA&\cite{ronneberger2015u}\cite{oktay2018attention}\cite{badrinarayanan2017segnet}\cite{diakogiannis2020resunet}&NA& 88 & Yes & No                 \\  \hline

\cite{liang20213d}&2021 &300(200)&NA&NA&\cite{cciccek20163d}&NA&89&No &No \\\hline

\cite{tian2021automatic} & 2021 & 70(50) &NA&NA&\cite{cciccek20163d}\cite{milletari2016v}\cite{zhang2019attention}\cite{huang20193d}\cite{isensee2021nnu}&NA& 94* & No & No                 \\\hline

\cite{pan2021coronary} & 2021 & 474(432)&NA&NA&\cite{yu20163d}&NA & 94 &No &No  \\\hline

\cite{cheung2021computationally}& 2021 & 69(44) & NA & NA & \cite{zhou2018unet++} & NA & 89 & No & No \\\bottomrule[2pt]

\hline

\multicolumn{10}{l}{* The segmentation result includes the initial part of the aorta. } \\
\multicolumn{10}{l}{$^{\dag}$ Two kinds of datasets including a private dataset and a public dataset are adopted for evaluation, and the comparison is only performed on}\\
\multicolumn{10}{l}{the public dataset.}\\
\multicolumn{10}{l}{$^{\ddag}$ The Dice score was estimated from figures. }\\
\end{tabular}



 
\label{tab:works}
\end{table*}

\section{Related Work}

In this section, we review existing literature on coronary artery segmentation.
We first review the related works by the type of approaches used, i.e.\ traditional ML-based approach 
and DL-based approach, 
and then discuss the datasets and benckmarks commonly adopted in these works.

\subsection{Traditional ML-Based Approach}

Hand-crafted features and prior knowledge of structures of coronary artery are extensively used in this approach.
\cite{zheng2011machine} proposed an ML method to exploit the rich domain-specific knowledge (particularly a set of geometric and image features) embedded in an expert-annotated dataset.
\cite{mohr2012accurate} proposed a level-set based approach for efficient processing.
\cite{wang2012vessel} combines level-sets with an implicit 3D model of the vessels for accurate segmentation.
\cite{broersen2012frenchcoast} adopted a pipeline consisting of three consecutive steps.
\cite{shahzad2013automatic} performed segmentation with the help of extracted centerlines.
\cite{lugauer2014improving} used  a learning-based boundary detector to enable a robust lumen contour detection via dense ray-casting.
\cite{lugauer2014precise} proposed a model-guided segmentation approach based on a Markov random field formulation with convex priors.
\cite{chi2015composite} integrated coronary artery features of density, local shape and global structure into a learning framework.
\cite{lesage2016adaptive} considered vessel segmentation as an iterative tracking process and proposed a new Bayesian tracking algorithm based on particle filters for the delineation of coronary arteries.
\cite{han2016automatic} used an active search method to find branches and seemingly disconnected but actually connected vessel segments.
\cite{freiman2017improving} used a flow simulation method \cite{nickisch2015learning} in the coronary trees with accounting for partial volume effects \cite{glover1980nonlinear}.
\cite{gao2019automatic} located the coronary root through extracting aorta by using circular Hough transform.
\cite{du2021automated} proposed a new segmentation framework which included noise reduction, candidate region detection, geometric feature extraction, and coronary artery tracking techniques.

\subsection{DL-Based Approach}
Since the rise of deep learning, it has attracted tremendous attention in the related communities.
Currently, there are mainly five technique trends including pixel-based segmentation \cite{moeskops2016deep,kjerland2017segmentation}, direct segmentation \cite{shen2019coronary,lee2019template,yang2019discriminative,fu2020mask,lei2020automated,gu2020segmentation,gu2021fusing,zhu2021segmentation,liang20213d,tian2021automatic,cheung2021computationally,li2018h,li20183d,lin2022calibrating}, patch based segmentation \cite{duan2018context,chen2018automatic,huang2018coronary,chen2019coronary,mirunalini2019segmentation,wang2021geometric,pan2021coronary}, tree data based segmentation \cite{kong2020learning}, and graph data based segmentation \cite{wolterink2019graph}.
Some other works like \cite{wu2019automated} performs detailed anatomical labeling of coronary arteries, which is out of the scope of this paper.

Pixel-based segmentation is the pioneer to adopt convolutional neural networks (CNNs) into coronary artery segmentation. 
\cite{moeskops2016deep} used a single convolutional neural network to show the feasibility of deep learning for coronary vessel segmentation.
\cite{kjerland2017segmentation} used two neural networks trained on aorta segmentation and coronary segmentation respectively and is able to segment the complete coronary artery tree.

Then, direct segmentation has been popular due to the rise of U-net \cite{ronneberger2015u}.
\cite{shen2019coronary} propose a joint framework based on deep learning and traditional level set method.
\cite{lee2019template} adopted template transformer networks where a shape template is deformed to match the underlying structure of interest through an end-to-end trained spatial transformer network.
\cite{yang2019discriminative} adopted a discriminative coronary artery tracking method which included two parts: a tracker and a discriminator.
\cite{fu2020mask} used Mask R-CNN for coronary artery segmentation in which the lung region prior is masked out to avoid the interferences from pulmonary vessels.
\cite{lei2020automated} integrated deep attention strategy into the fully convolutional network (FCN) model \cite{long2015fully} to highlight the informative semantic features extracted from CCTA images. 
\cite{gu2020segmentation} proposed a global feature embedded network for better coronary arteries segmentation.
\cite{gu2021fusing} proposed a two-stage strategy that retains the advantages of both 2D and 3D CNNs.
\cite{zhu2021segmentation} proposed a U-shaped network based on spatio-temporal feature fusion structure which combines features of multiple levels and different receptive fields separately to get more precise boundaries.
\cite{liang20213d} proposed an improved U-net which integrated channel attention and spatial attention to distinguish confusing categories and targets with similar appearance features. 
\cite{tian2021automatic}  combined deep learning and digital image processing algorithms to address the problem of the limited GPU memory.
\cite{cheung2021computationally} proposed a fully automatic two-dimensional U-net model to segment the
aorta and coronary arteries.

As direct segmentation usually requires a downsampled image as the input thus cannot obtain the local details well, patch based segmentation becomes popular. Essentially in this method, a rough segmentation on the downsampled image is performed first and then a refinement is adopted in a patch manner to extract the local details.
\cite{duan2018context} proposed a context aware 3D FCN for vessel enhancement and segmentation which outperformed conventional Hessian vesselness based approach \cite{frangi1998multiscale}.
\cite{chen2018automatic} adopted a paired multi-scale 3D CNN to identify which voxels belong to the vessel lumen.
\cite{huang2018coronary} transformed CTA image into small patches and then sent them to a 3D U-net \cite{cciccek20163d} for processing.
\cite{chen2019coronary} incorporate vesselness maps into the input of 3D U-Net \cite{cciccek20163d} to highlight the tubular structure of coronary arteries.
\cite{mirunalini2019segmentation} combined CNN and recurrent neural networks to identify the presence of coronary arteries in 2D slices. 
\cite{wang2021geometric} incorporated the voxel and point cloud-based segmentation methods into a coarse-to-fine framework.
\cite{pan2021coronary} proposed 3D Dense‐U‐Net which was further optimized with a focal loss to tackle the imbalance problem.

Recently, some methods have been proposed to incorporate image data into segmentation by converting them into special data structures, such as tree structures and graph structures.
Note that the morphological structure of the coronary arteries is tree-like, and generally divides into a left coronary artery and a right coronary tree with blood flowing from the aorta to the coronary arteries and then to the individual branches.
\cite{wolterink2019graph} used graph convolutional networks (GCNs) to predict the spatial location of vertices in a tubular surface mesh that segments coronary artery. 
\cite{kong2020learning} proposed a novel tree-structured convolutional gated recurrent unit (ConvGRU) model to learn the anatomical structure of coronary artery.

\subsection{Benchmark and Datasets}
A more detailed analysis of existing works are shown in Table \ref{tab:works}, in which we can easily notice the broad attention of this topic around the world.
We can also find that most works did not make a fair and comprehensive comparison with earlier works.
\textcolor{ForestGreen}{For the traditional ML-based approach, many works focused on algorithmic development based on a public dataset \cite{kiricsli2013standardized} which has only 42 CTA images.}
The comparisons were also made based on this dataset, and each work \cite{mohr2012accurate,wang2012vessel} just obtained the performance from other papers without reimplementation.
In the meantime, other works \cite{han2016automatic} were using another public dataset \cite{schaap2009standardized} for their evaluation.
On the other hand, however, even more works prepared in-house datasets for evaluation which failed to compare 
to other works or only compared with general methods \cite{lugauer2014precise,chi2015composite,han2016automatic,freiman2017improving,gao2019automatic,du2021automated}.
Note that we define general methods here as those usually target at general algorithms for a variety of applications, while oppositely specific methods as thoses designed and optimized for specific applications, e.g., coronary vessel segmentation.
The use of proprietary dataset and hence missing or partial comparison problem became even more common when 
DL-Based approach began to prevail.
For example, \cite{kong2020learning} did not compare their model with any previous related works but just some popular DNNs such as 3D U-net \cite{huang20193d}.
\cite{shen2019coronary} listed the Dice score achieved by other works on different dataset for the evaluation of theirs.
In addition to dataset inconsistency, problems in other benchmarking components are also widespread.
Some works \cite{chi2015composite,han2016automatic} used different sets of evaluation metrics based on their specific clinical needs, while others \cite{shen2019coronary} used a different annotation strategy in which the initial aorta is included, leading to a much higher Dice score. 


\begin{figure*}[h]
\centerline{\includegraphics[width=1\textwidth]{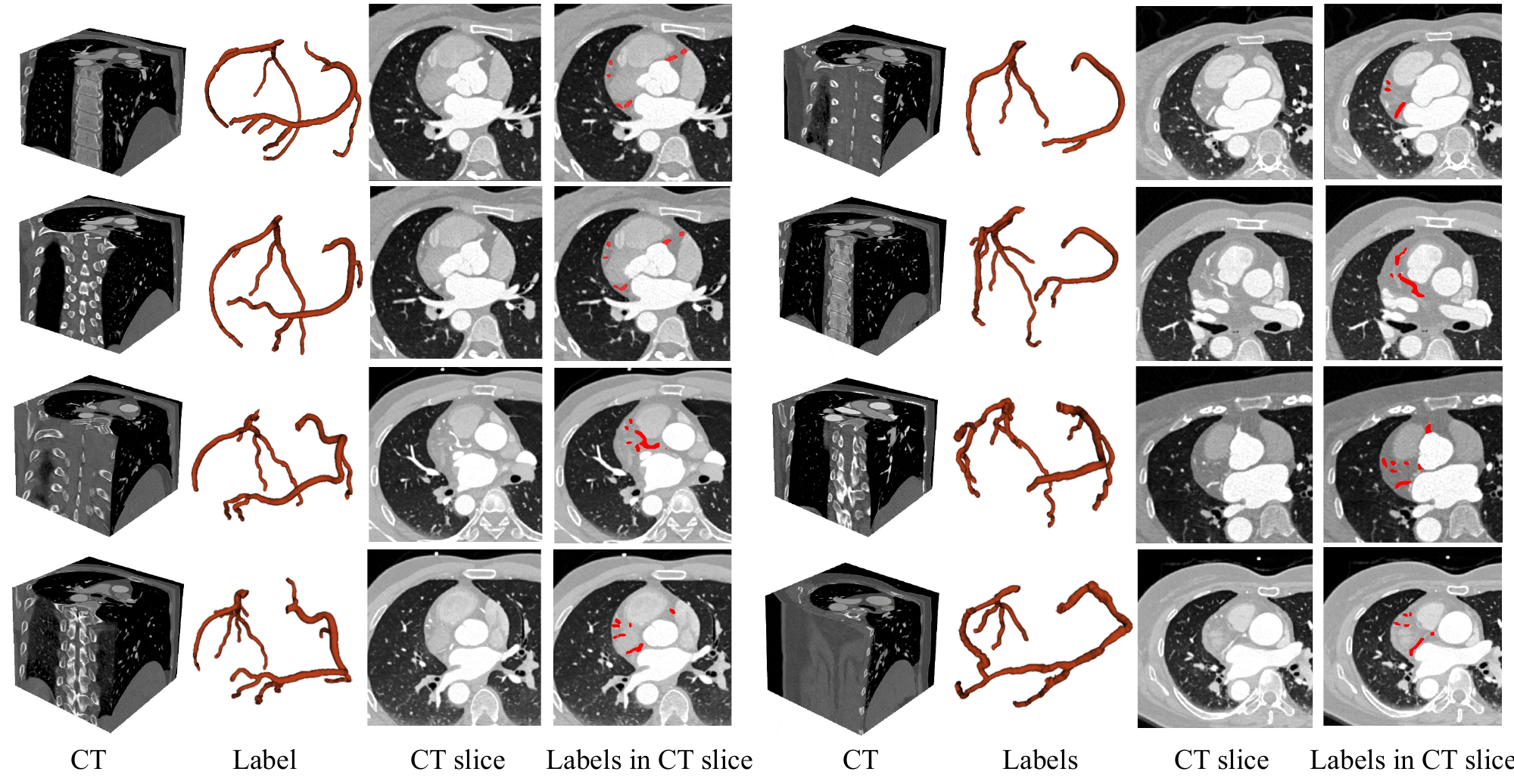}}
\caption{\textcolor{ForestGreen}{Examples including CT images, their labels, CT slices and their labels in CT slices in the proposed ImageCAS dataset. Note that subclasses of coronary arteries are not further individually labeled.}}
\label{fig_dataset}
\end{figure*}

\begin{figure*}[hbt!]
\centerline
{\includegraphics[width=1\textwidth]{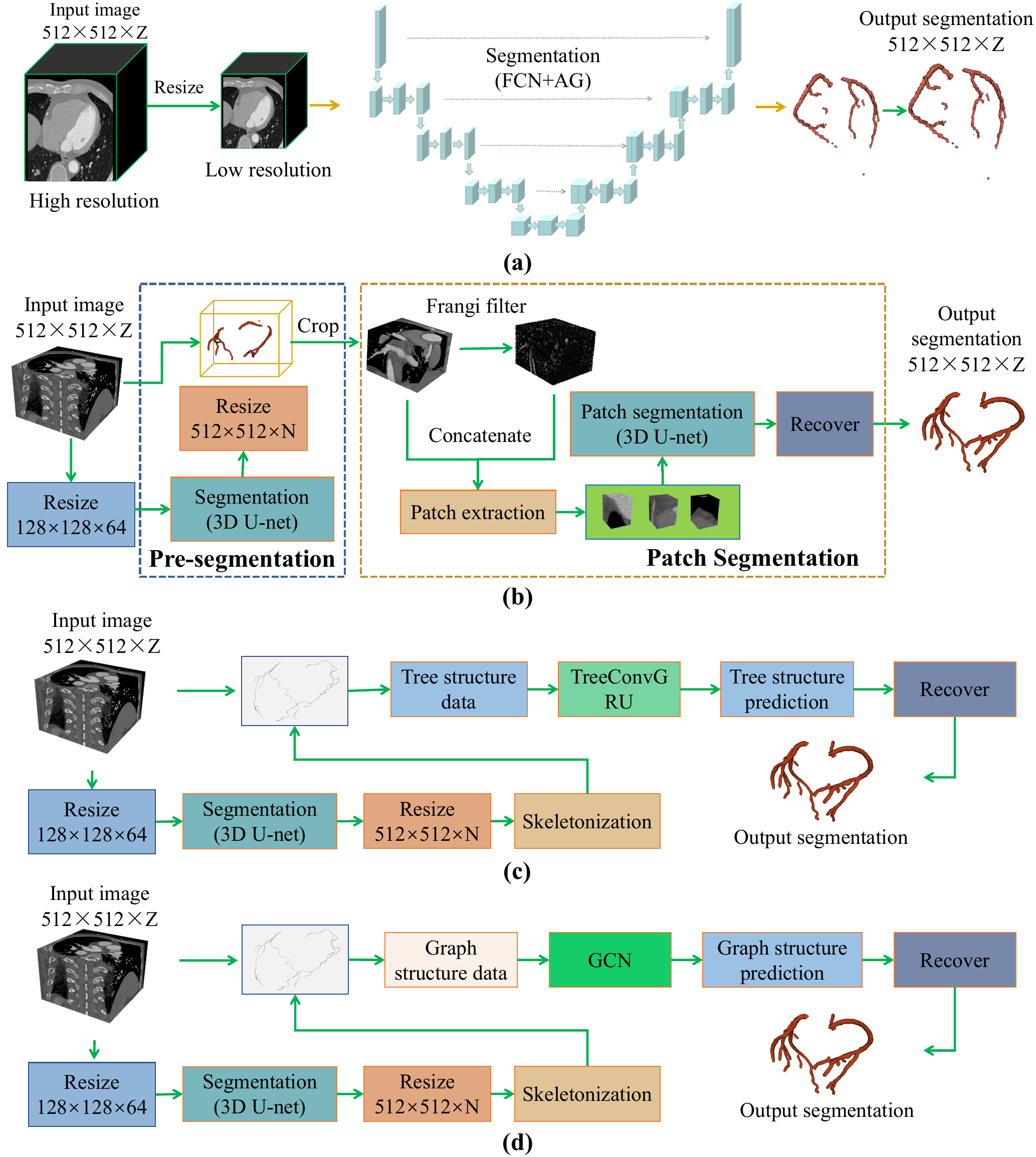}}
\caption{Overview of the methods in the proposed benchmark including (a) direct segmentation \cite{shen2019coronary}, (b) patch based segmentation \cite{huang2018coronary,chen2019coronary}, (c) tree data based segmentation \cite{kong2020learning}, and (d) graph data based segmentation \cite{wolterink2019graph}.}
\label{fig_existing_method}
\end{figure*}

With the above analysis, we can notice that a benchmark and a dataset for further development of this topic are needed.
On one hand, most of the methods only used in-house datasets for evaluation, but their image acquisition parameters, reconstruction techniques, and label methods vary significantly. 
On the other hand, most of the methods just performed comparison with general methods but without existing related works.
The situations became even worse for deep learning based methods.
Note that the success of deep learning heavily relies on a large dataset with high quality annotations, but the two commonly used datasets \cite{schaap2009standardized,kiricsli2013standardized} are quite small (only contain just 8 and 18 images respectively for training).
In addition, almost all works have not provided their source code, which also brings a lot of difficulties for their fair comparison.
Although some works \cite{moeskops2016deep,kjerland2017segmentation,yu2017automatic} are relatively easy to implement for comparison purpose, most others are intensively complex with too many details in their specifications such as network structure, preprocessing, hyper-parameters, and postprocessing.
The above problem has also been identified in some other works.
For example, the work \cite{tian2021automatic} stated that "since most of the coronary artery segmentation methods use private dataset, there is no suitable public coronary data set for us to test."
Both \cite{huang2018coronary} and \cite{yang2019discriminative} performed comparison with related works on the dataset \cite{schaap2009rotterdam}, but conducted experiment on a private dataset without comparison which is due the fact that it is not easy to implement the existing methods.
The work \cite{huang2018coronary} stated that "the training data seemed relatively small...".
Some works \cite{yang2019discriminative} also tried to solve the small dataset problem.
The work \cite{moeskops2016deep} stated that "in future work, we will further investigate the capacity of the current architecture with more data and segmentation tasks.".

In order to uniformly investigate the effectiveness of each method, we collected a large dataset and made it public available.
This dataset contains 1000 3D CTA images, which is considerably larger than the existing public datasets.
Furthermore, we proposed a benchmark in which we tried our best to implement typical existing methods. 
Although the code of the existing methods is not publicly available and some experimental details are missing, we have conducted similar implementations based on their core ideas.
We further evaluated the performance of each method on the proposed dataset with multiple sets of configurations.
In addition, we have also proposed a baseline segmentation framework which achieved better performance than existing works. 
Details of the proposed dataset and benckmark will be discussed in Section \ref{sec:dataset} and \ref{sec:benchmark}.

\section{ImageCAS Dataset}\label{sec:dataset}
The proposed dataset consists of 3D CTA images captured by Siemens 128-slice dual-source scanner from 1000 patients.
For those patients who have previously been diagnosed with coronary artery disease, early revascularization (within 90 days after) are included.
The high-dose CTA is performed and during the reconstruction, the 30\%-40\% phase or the 60\%-70\% phase is selected to obtain the best coronary artery images. 
The images have sizes of $512\times 512 \times (206-275)$ voxels, a planar resolution of 0.29-0.43 $mm^{2}$, and spacing of 0.25-0.45 $mm$. 
The data was collected from realistic clinical cases at the Guangdong Provincial People's Hospital during April 2012 to December 2018. Only the patients older than 18 years and with a documented medical history of ischemic stroke, transient ischemic attack and/or peripheral artery disease are eligible  to be included. 
The index cardiac CTA or low imaging quality of CCTA (assessed by level III radiologist) were also excluded due to the possible influence on the function of the coronary artery.
Finally, there are totally 414 females and 586 males included, the average ages being 59.98 and 57.68, respectively. 
The left and right coronary arteries in each image are independently labeled by two radiologists, and their results are cross-validated. 
In case of discrepancy, a third radiologist will perform the annotation and the final result is determined by consensus. 
The labeled coronary artery includes the left main coronary artery, left anterior descending coronary artery, left circumflex coronary artery, right coronary artery, diagonal 1, diagonal 2, diagonal 3, obtuse marginal branch 1, obtuse marginal branch 2, obtuse marginal branch 3, ramus intermedius, posterior descending arteries, 
acute marginal 1 and other blood vessels according to the AHA naming convention (17 paragraphs).
Two examples of the dataset are shown in Fig. \ref{fig_dataset}.

\section{Benchmark}\label{sec:benchmark}
As the deep learning based methods in almost all recent works have shown promising performance in cornary artery segmentation,
we have implemented several typical deep learning based methods including direct segmentation \cite{shen2019coronary}, patch based segmentation \cite{huang2018coronary,chen2019coronary},
tree data based segmentation \cite{kong2020learning}, graph data based segmentation \cite{wolterink2019graph}.
In addition, we also propose a baseline method which achieves better performance than existing works on the proposed
large-scale dataset.
We describe the details of each method in the following sections and more about the their implementations can be found at
https://github.com/XiaoweiXu/ImageCAS-A-Large-Scale-Dataset-and-Benchmark-for-Coronary-Artery-Segmentation-based-on-CT.


\subsection{Direct Segmentation}
Direct segmentation only adopts a single neural network, which is a simple but efficient for coronary artery segmentation.
Particularly, the image is fed to the network which outputs the probability map without complex processing.
As the images are usually large and thus hard to be accommodated by GPUs, they need to be downsized to a smaller resolution in practice.
We select FCN-AG \cite{shen2019coronary} as a representative.
The details of the method are shown in Fig. \ref{fig_existing_method}(A), and the overall structure is an FCN backbone with attention gate modules.
The process can be divided into the following three steps:
\begin{enumerate}
    \item{resize high resolution CTA images into low resolution ones using interpolation;}
    \item{feed the low resolution image to the FCN-AG network, and then obtain the prediction of coronary artery.}
    \item{resize low resolution prediction labels into the high resolution ones using interpolation.}
\end{enumerate}
For more details of the method, please refer to the original work \cite{shen2019coronary}.

\subsection{Patch based Segmentation}


To tackle the computational resource constraints and missing details due to downsampling, patch based segmentation \cite{huang2018coronary,chen2019coronary}  has been proposed.
We consider the dual-CNN-based framework \cite{huang2018coronary,chen2019coronary} as an example to evaluate this class of methods (Fig.\ref{fig_existing_method}(b)).
First, a 3D U-net is used to extract the region of interest (RoI) and remove irrelevant areas.
Then, the cutout image is resized from $512 \times 512 \times Z$ to $W \times H \times Z_c$ ($W \leq512,H \leq512, Z \leq Z_c$).
According to the \cite{chen2019coronary}, we use Frangi filtering to enhance tubular structures, and the obtained vascular enhancement map is combined with the input image to form a multi-channel input.
Next, the multi-channel input is decomposed into small patches which are fed into another CNN for processing.
Finally, the segmentation of the patches are combined to obtain the final segmentation result.

\subsection{Tree Data Based Segmentation}


Tree data based segmentation method is promising as it considers the morphological structure of the coronary arteries.
We select the work presented in \cite{kong2020learning} to implement a tree convolutional recurrent neural network and details are shown in Fig. \ref{fig_existing_method}(c).
First, the approximate centerline is obtained by skeletonizing the pre-segmented labels.
To simplify the steps of constructing the tree, the point with the largest coordinate in the Z-axis is taken as the root node, and the rest are regarded as the leaf nodes.
Then, each centerline point is taken as the center to get the patch which is then used to extract node features.
Next, the connections in the tree are constructed based on the neighborhood relationship between centerline points.
Finally, the tree structure data is used as the input to TreeConvGRU \cite{kong2020learning} and the predicted labels are obtained as output.

\subsection{Graph Based Segmentation}
Graph based segmentation works in a similar way as tree data based segmentation.
Referring to the idea in \cite{wolterink2019graph}, we have designed a similar scheme as shown in Fig. \ref{fig_existing_method}(d).
First, the centerline is obtained by the same processing pipeline as that in tree data based segmentation.
Then, graph structured data is produced.
Particularly, a number of rays perpendicular to the tangent of each centerline point are issued while the angle formed by the two tangents between adjacent rays remains the same.
A d-dimensional feature is formed from the starting point of the ray in the direction of the tangent line at some step size, depending on the block of voxels through which the ray passes.
Each ray intersects the edge of the vessel, and the Euclidean distance between the point of intersection and the centerline point is obtained as its radius.
After predicting the radii of all the centerline point, the coronary arteries are reconstructed and we can get the final segmentation.

\subsection{Baseline Method}
\begin{figure*}[h]
\centerline
{\includegraphics[width=1\textwidth]{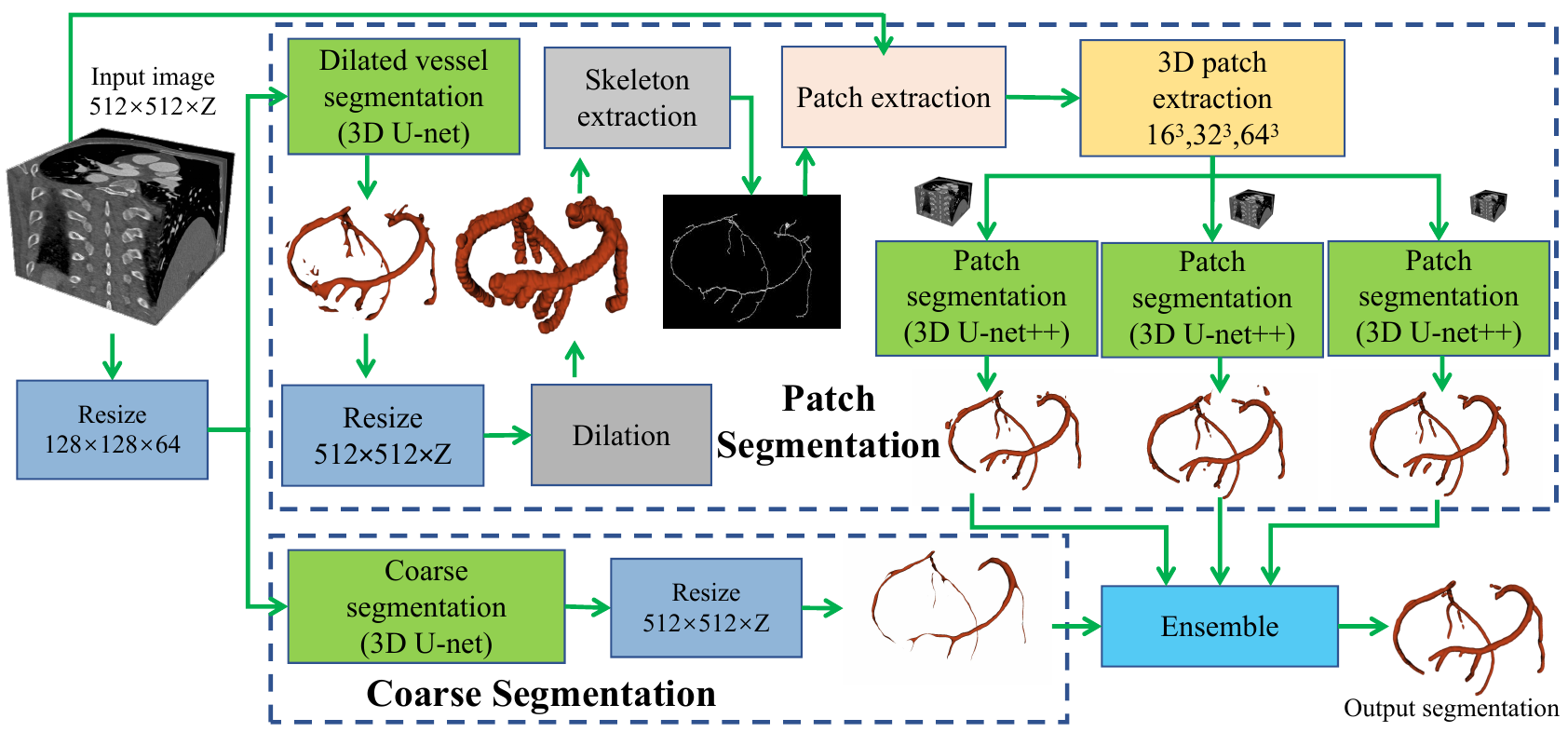}}
\caption{Framework of the proposed baseline method.}
\label{fig_pipe}
\end{figure*}
The proposed baseline method is a combination of patch segmentation and coarse segmentation as shown in Fig. \ref{fig_pipe}.
Such combination is a trade-off between performance and implementation feasibility.
On one hand, the direct segmentation of the whole 3D image in the original resolution is not feasible due to the large memory consumption, while the segmentation of the resized image (defined as coarse segmentation) is feasible but leading to a limited performance. 
On the other hand, patch based segmentation can provide more details but sometimes also has obvious errors due to the loss of global contextual information.

Our baseline method has two main modules: patch segmentation and coarse segmentation.
Before being processed by the two modules, each input image is resized from $512\times 512 \times (206$-$275)$ to $128\times 128 \times 64$ using linear interpolation.
In coarse segmentation, a 3D U-net \cite{cciccek20163d} is used to obtain a rough segmentation of coronary artery which can usually capture the global structure, e.g., all the related vessels are included with inaccurate boundaries. 
For the ease of discussion, we denote $\theta(X)$ as the segmentation network, where $X$ is the input image, and $\hat{Y_{c}}$ is the output.
Then, the network is trained with the following similarity coefficient loss:
\begin{equation}\label{eq1}
	L_{c}(\theta(X),Y)=1-\frac{2\mid\hat{Y_{c}}\cap Y\mid+s}{\mid \hat{Y_{c}} \mid+\mid Y\mid +s},
\end{equation}
where $Y$ is the ground truth and $s$ is the smoothing factor.

In patch segmentation, the main problem is how to precisely obtain the relevant patches which contains coronary arteries. 
If we simply use sliding window to generate patches, all the patches including the ones that contain other small vessels or vessel-like anatomies will be processed.
In this way, the segmentation network needs to consider both the target regions and the non-target regions, which introduces much difficulty into the segmentation task.
To solve this problem, we adopt another network to roughly obtain the overall target region.
Note that the result in the coarse segmentation step can be adopted for such purpose.
However, in the prior experiments we find that this can lead to a number of problems, such as interrupted vessels.
Thus, we introduce another sub-step, \textit{dilated vessel segmentation} to extract a rough mask of the coronary artery using 3D U-net.
Then, the output is dilated, and the vessels become much thicker and less likely to be interrupted.
To make the vessels even thicker, the following weighted similarity coefficient loss function \cite{sudre2017generalised} is used:
\begin{equation}\label{eq2}
	L_{m}(\phi(X),Y)=1-\frac{\mid \hat{Y_{d}}\cap Y\mid+s}{\alpha\mid \hat{Y_{d}} \mid+(1-\alpha)\mid Y\mid +s},
\end{equation}
where $\phi(X)$ is the segmentation network, $\hat{Y_{d}}$ is the output, $s$ is the smoothing factor, and $\alpha$ is the class weight ($\alpha \in (0,1)$).
In order to make the segmentation results to cover most of target regions, $\alpha$ is set to 0.01. 
Thus, the network is biased to obtain oversized vessels in the output. 
In the network training, $\hat{Y_{d}}$ is obtained by dilating the coronary artery in the ground truth $Y$. 
To obtain the precise position, $\hat{Y_{d}}$ is further resized to the original size of the input image using interpolation.
\textcolor{ForestGreen}{Note that during the test stage, the results from the dilated vessel segmentation network are further dilated to insure better connectivity.}
The skeleton of the vessels is extracted using a surface thinning algorithm \cite{lee1994building}.

With the resized skeleton and the input image, we perform patch segmentation as follows:
(1) Extracting the two largest connected components through connected component analysis and discarded others. This is due to the domain knowledge that there are usually two vessels (the left and right coronary arteries); 
(2) Extracting $nc$ sets of cubic patches with a skeleton point as the center and an edge length of $r$ ($nc=3$, and $r=16, 32, 64$ in Fig. \ref{fig_pipe});
(3) 3D U-net++ \cite{zhou2018unet++} is adopted to process $nc$ sets of patches, as it handles details more accurately than 3D U-net;
(4) The segmented patches are then fused to obtain a segmentation image of the same size as the original input image;
(5) Finally, the $nc$ segmentation images are fed to the ensemble step to get the final output. 
\textcolor{ForestGreen}{Note that  majority voting is adopted here for the ensemble.}


\section{Experiments and Discussion}

In this section, we first discuss the overall setup for all experiments.
Then, each method with specific configurations and their performances are discussed. Finally, all the methods with the optimal configurations are compared and analyzed.  

\subsection{Experiment setup}
All experiments were implemented using PyTorch \cite{paszke2019pytorch} and DGL \cite{wang2019deep}, and performed on a Nvidia RTX 3090 GPU with a 24G memory. 
In the direct segmentation, patch segmentation and tree-structure segmentation, we used the Dice loss during training.
In the graph based segmentation, distance values are cubed in the loss function as in \cite{wolterink2019graph}, and we adopted the loss functions as shown in Eq. 1 and Eq. 2 for training. 
In patch based segmentation, the patches with 16,32,64 are discussed, and the Dice loss is used for training. 
In the baseline method, a spherical structure with a radius of $r=5$ is used to dilate the vessels.
All the networks in our implementation are trained for 30 epochs (\textcolor{ForestGreen}{about 21,000 iterations}), and the Adam optimizer is adopted with a learning rate of 0.002. 
\textcolor{ForestGreen}{Due to the limited GPU memory, the batch size various for different input sizes.
The batch size is 8, 2, and 1 for an input size of $128\times128\times128$, $256\times256\times128$, and  $512\times512\times256$, respectively in the pre-segmentation or coarse segmentation steps.
While that in the patch segmentation step is 512, 64, 8 for an input size of $16^3$, $32^3$, and $64^3$, respectively.
}
Experiments were evaluated using a 4-fold cross-validation approach, with a training set of 750 cases (50 cases are used for validation) and a test set of 250 cases.
The Dice score is used for evaluation, which is widely used in the community as indicated in Table \ref{tab:works}.


\subsection{Configuration discussion}

\begin{figure*}[htb]
\centerline{\includegraphics[width=0.9\textwidth]{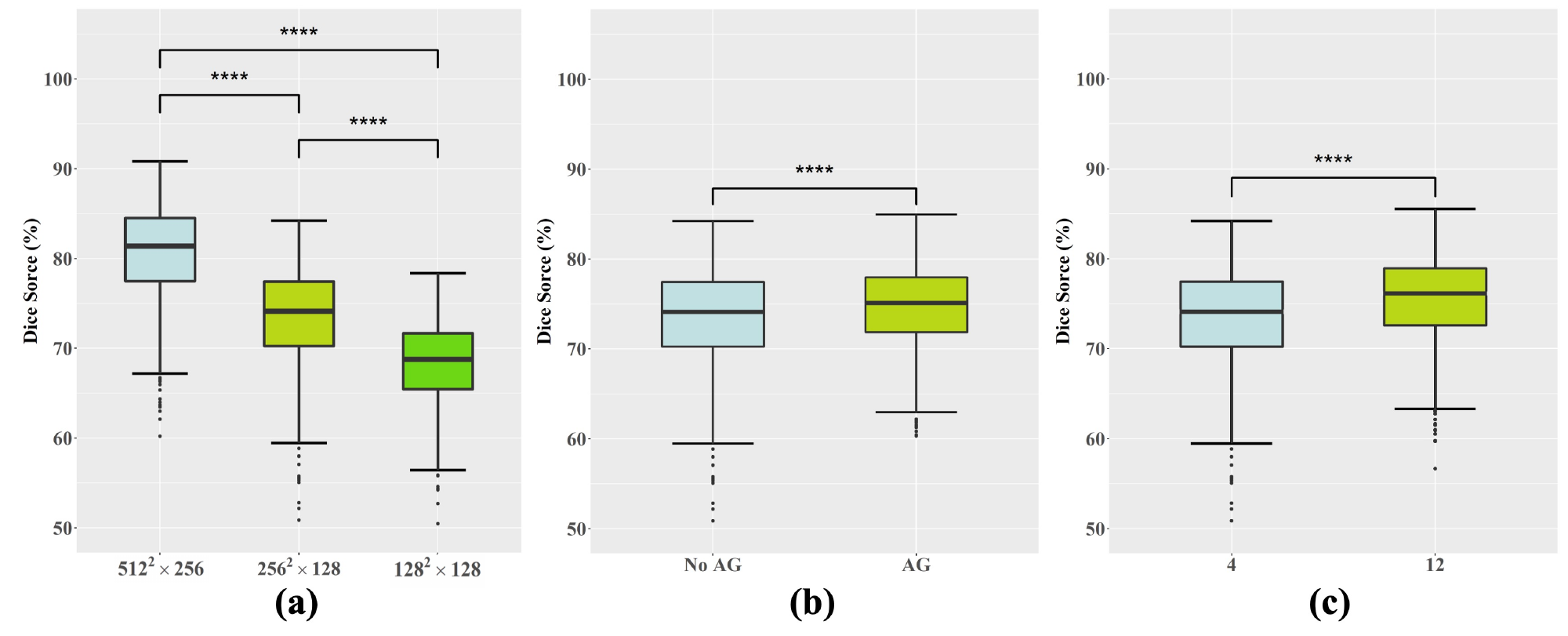}}
\caption{Performance discussion of direct segmentation with various configurations including (a) input size, (b) attention gate module, and (c) the number of channels. $ns$ stands for not significant ($p$$>$0.05), and ** stands for $p$ smaller than 0.01, *** stands for $p$ smaller than 0.001, and **** stands for $p$ smaller than 0.0001.}
\label{fig_direct_seg_result_quantity}
\end{figure*}

\begin{figure*}[htb]
\centerline{\includegraphics[width=1\textwidth]{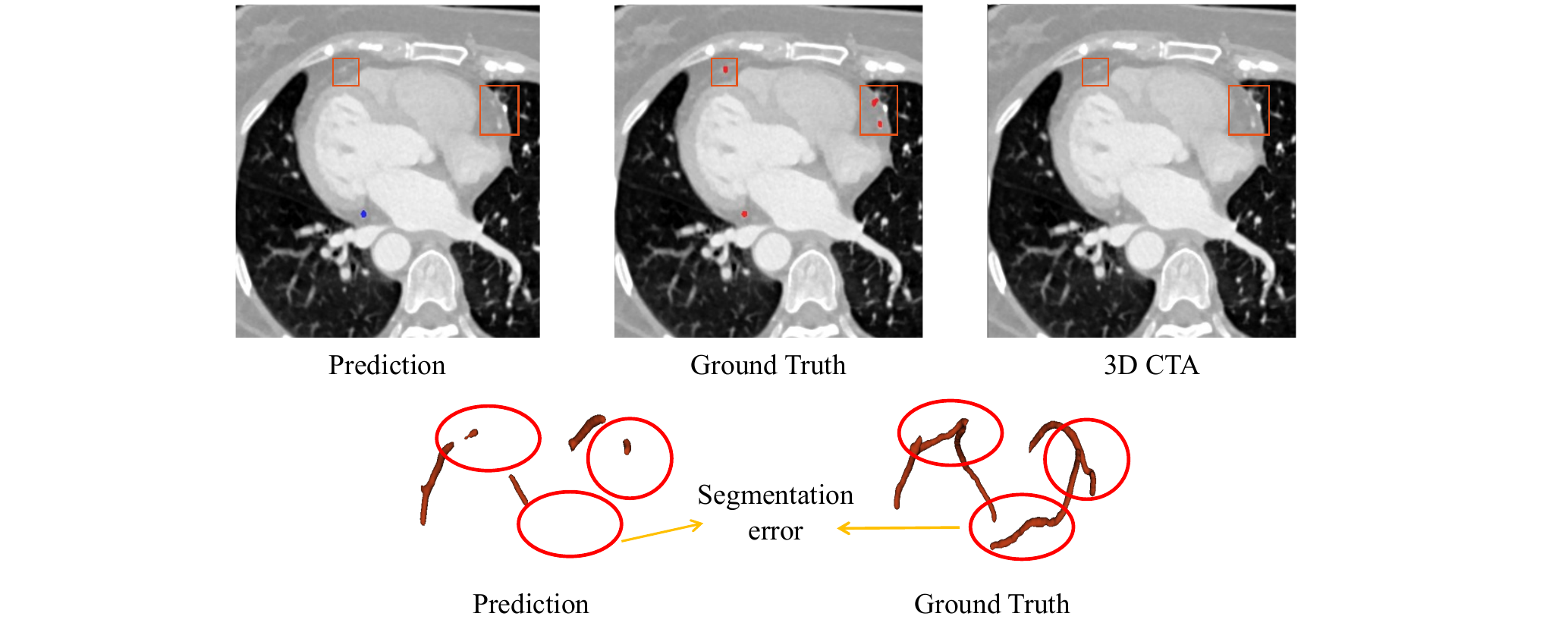}}
\caption{\textcolor{ForestGreen}{Visual discussion of direct segmentation results with failure cases.}
}
\label{fig_direct_seg_result_visual}
\end{figure*}

\subsubsection{Direct segmentation}
We discussed some factors that may affect the segmentation performance including the input size, the use of attention mechanism and the number of channels.
Input size including $128\times128\times128$, $256\times256\times128$, and  $512\times512\times256$ are discussed which are obtained by nearest neighbour interpolation of the original image.
The number of channels including 4 and 12 are discussed.

The results are shown in Fig. \ref{fig_direct_seg_result_quantity}.
We can notice that the input size of $512\times512\times256$ significantly improves the Dice score by 7.38\% (p\textless0.0001) and 12.32\% (p\textless0.0001) compared to that of $256\times256\times128$ and $128\times128\times128$, respectively.
The addition of the attention gate module significantly improves the performance by 1.34\% (p\textless0.0001), which is also observed by the work \cite{shen2019coronary}.
The number of channel of 12 (parameter number of 5.24M) obtains a significantly better Dice score than that of 4 (parameter number of 0.59M) by 2.13\% (p\textless0.0001) in Dice score.

Visual discussion is shown in Fig. \ref{fig_direct_seg_result_visual}.
Due to the low contrast around the coronary region in the CTA images, direct segmentation always focus on the whole coronary artery while ignoring the local details.
We can notice that there is little difference in voxel intensity between the vascular portion and the rest of the adjacent tissue, and the direct segmentation method cannot correctly identify this portion of coronary artery resulting in segmentation errors.
In addition, directly segmentation with a high resolution input needs more computational resources and limits the network size and the model capability.

\subsubsection{Patch based segmentation}


In the implementation, the cropping method in \cite{chen2019coronary} was adopted, where real labels are used in the training set for skeletonization. 
The corresponding patches are cropped at the centre of the skeleton points, and the cropped regions are randomly selected so that the ratio of regions with labels to those without is 1:1.
We explored three factors including patch size, frangi channel and data augmentation that may affect the performance.
For data augmentation, the probability of rotation (randomly 0, 90, 180, 270 degrees) and horizontal flipping are discussed.
For ease of discussion, the two probabilities are set to the same value, and three values including 0, 0.2, and 0.5 are discussed.

\begin{figure*}[!htb]
\centerline{\includegraphics[width=0.91\textwidth]{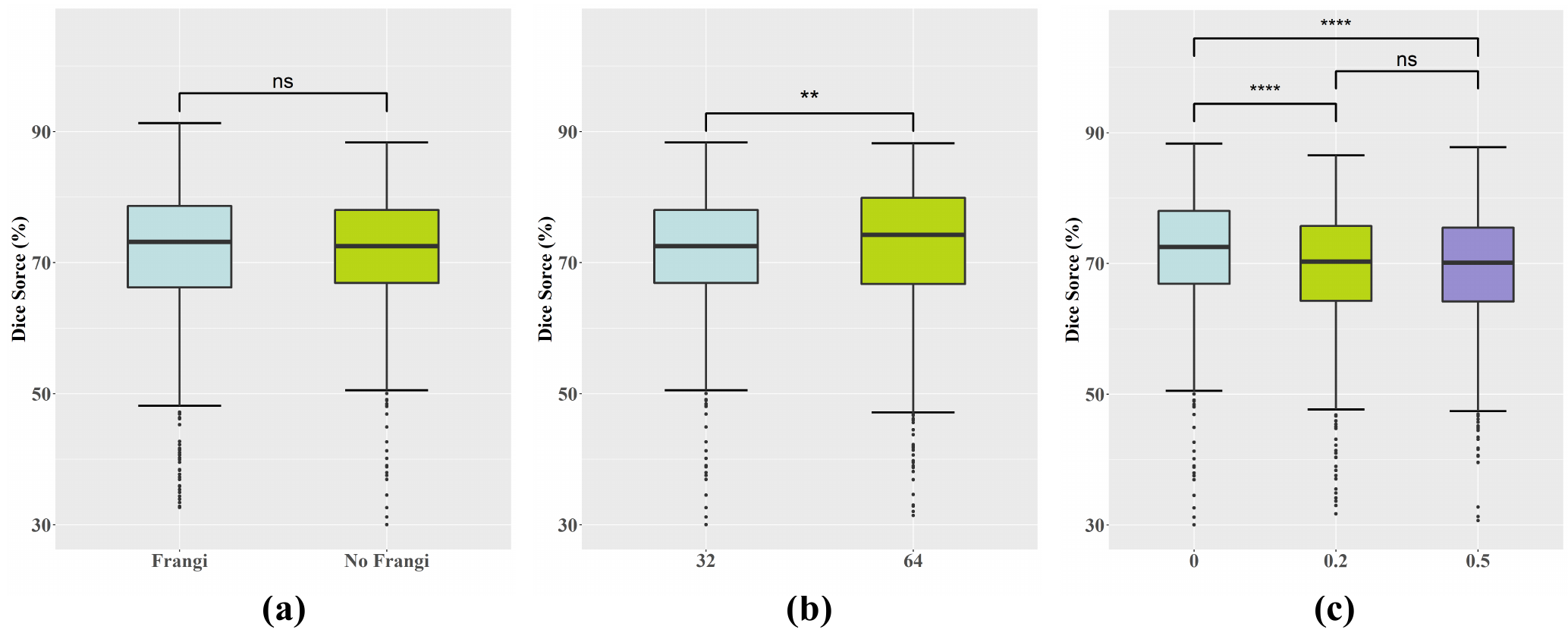}}
\caption{
Performance discussion of patch based segmentation with various configurations including (a) Frangi channel, (b) patch size, and (c) data augmentation. $ns$ stands for not significant ($p$$>$0.05), and ** stands for $p$ smaller than 0.01, *** stands for $p$ smaller than 0.001, and **** stands for $p$ smaller than 0.0001.}
\label{fig_patch_seg_result_quantity}
\end{figure*}

Quantitative performance is shown in Fig. \ref{fig_patch_seg_result_quantity}.
There is no significant difference between the network with the Frangi channel and that without, with a difference of only 0.01\% (p\textgreater0.05) in the Dice score.
For patch size, we can notice that a larger patch size obtains a significantly higher Dice score than a smaller one, which is expected as a larger patch size has a larger receptive field and thus can capture better context information.
For data augmentation, we find that flipping and rotation probabilities of 0/0 (no flipping and rotation) obtains an improvement of 2.63\% (p\textless0.0001) and 2.73\% (p\textless0.0001) in Dice scores than that of 0.2/0.2 and 0.5/0.5.
This interesting phenomenon may due to the fact that coronary arteries have its own directions with corresponding surround anatomies, and rotation and flipping operation may produce unrealistic augmented samples which may harm the training process.

\begin{figure*}[!htb]
\centerline{\includegraphics[width=0.95\textwidth]{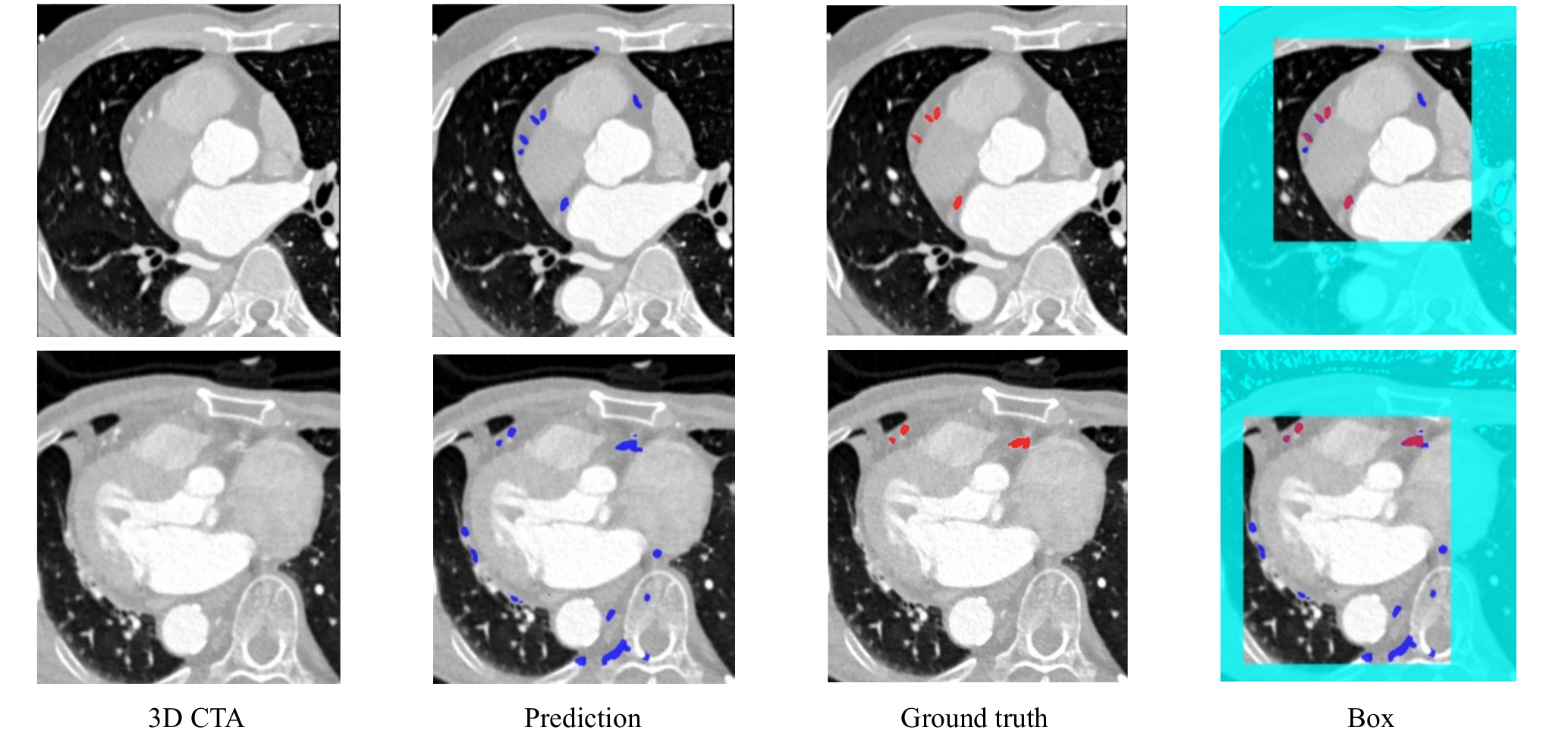}}
\caption{Visual discussion of failed cropped RoI in patch based segmentation including cross-sectional view on the input images, predicted labels, ground truths and pre-segmented formed bounding boxes.}
\label{fig_patch_seg_result_visual}
\end{figure*}

\begin{figure*}[!htb]
\centerline{\includegraphics[width=1\textwidth]{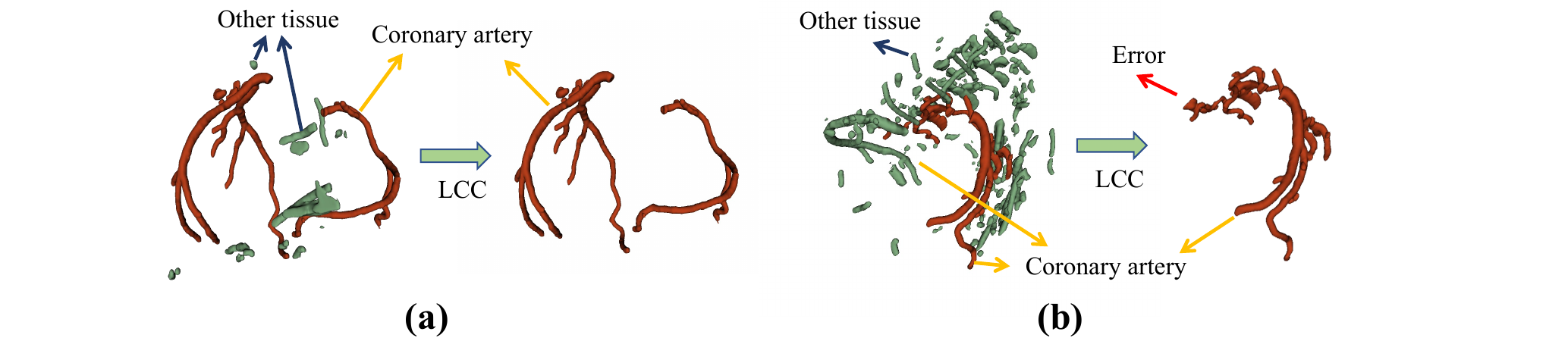}}
\caption{Visual discussion of (a) successful and (b) failed cases of largest connected component (LCC) post-processing in patch based segmentation.}
\label{fig_patch_seg_result_visual2}
\end{figure*}

Visual discussion of failed cropped regions of interest (RoI) in patch based segmentation is shown in Fig. \ref{fig_patch_seg_result_visual}.
We can see that several similar vessels are recognized as coronary arteries which may caused by the fact that their appearance are quite similar to the coronary arteries.
To tackle this problem, we can adopt a connectivity domain analysis in the post-processing step, and we find that for most of the predicted images, this refinement is effective in removing similar regions.
However, this does not work well for some images, generating large number of small vessels and resulting in the partial removal of the coronary arteries as shown in Fig. \ref{fig_patch_seg_result_visual2}.
In Fig. \ref{fig_patch_seg_result_visual2}(a), the border is well removed from the tissue around the heart, while in Fig. \ref{fig_patch_seg_result_visual2}(b) the pre-segmentation holds other structures such as bone tissue in the output.

\subsubsection{Tree data based segmentation and graph based based segmentation}
\begin{figure*}[htb]
\centerline{\includegraphics[width=1\textwidth]{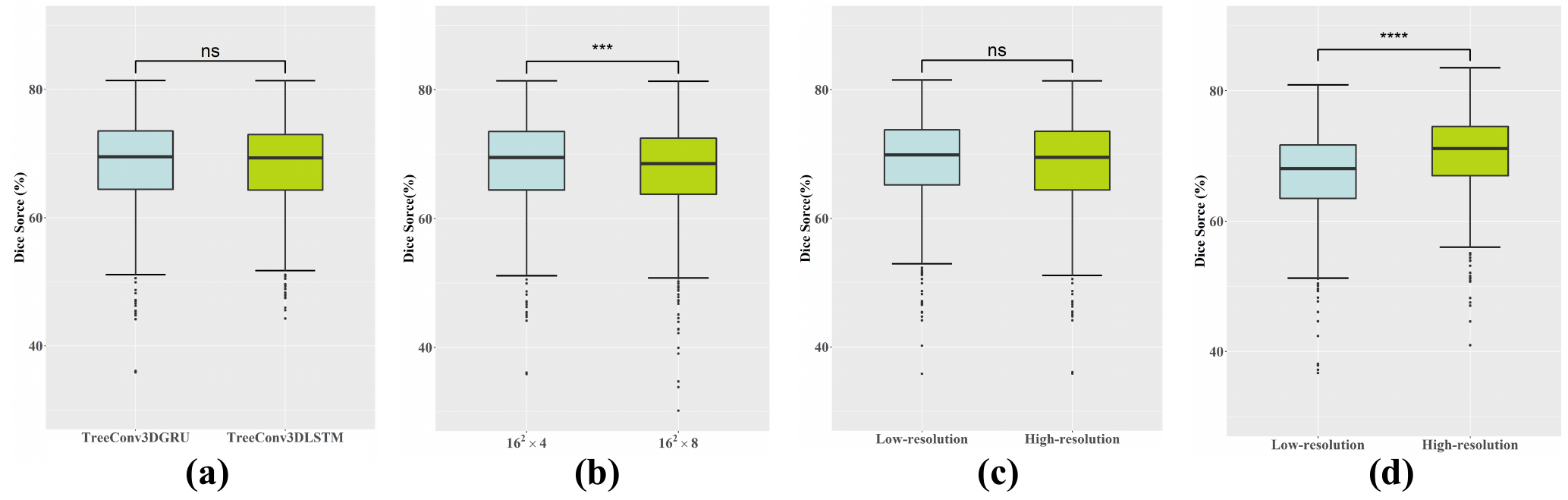}}
\caption{Performance discussion of tree data based segmentation and graph based segmentation with various configurations including (a) tree models, (b) patch size in tree node construction, input size in (c) tree data based segmentation and (d) graph based segmentation. $ns$ stands for not significant ($p$$>$0.05), and ** stands for $p$ smaller than 0.01, *** stands for $p$ smaller than 0.001, and **** stands for $p$ smaller than 0.0001..
}
\label{fig_tree_graph_seg_result_quantity}
\end{figure*}

\begin{figure*}[htb]
\centerline{\includegraphics[width=1\textwidth]{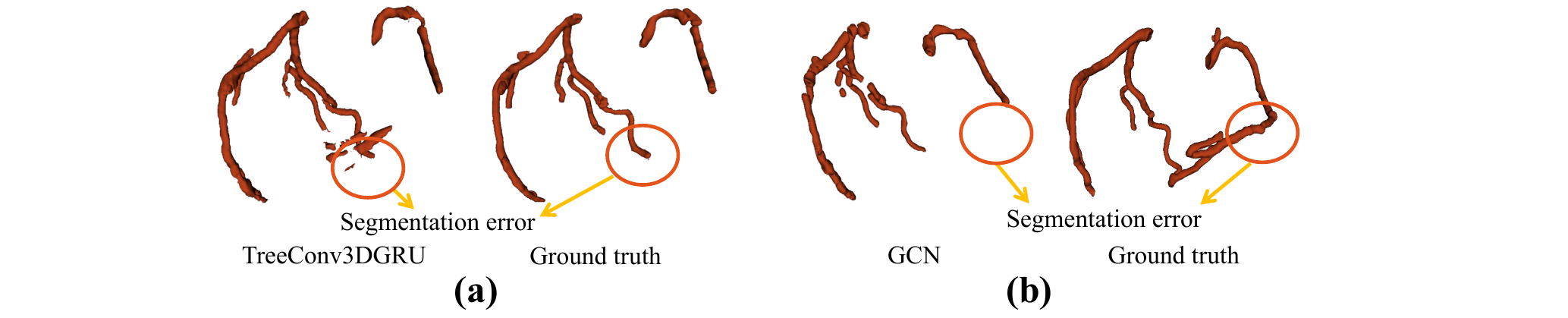}}
\caption{Visual discussion of failed cases in tree and graph based segmentation.
}
\label{fig_tree_graph_seg_result_visual}
\end{figure*}
As tree data based segmentation and graph based based segmentation are similar and share the same pre-segmentation module, we put the two together for ease of discussion.
In both the two methods, the input size including $128\times128\times128$ and $512\times512\times256$ are discussed.
In the tree structure segmentation, we adopted two models including TreeConv3FGRU and TreeConv3DLSTM and the patch size including $16\times16\times4$ and $16\times16\times8$ during tree node construction for discussion. 

Segmentation performance is shown in Fig. \ref{fig_tree_graph_seg_result_quantity}.
We can discover that the two tree models TreeConvGRU and TreeConvLSTM has only a difference of 0.06\% (p\textgreater0.05) in Dice score thus with no significant difference.
In tree data based segmentation, the patch size of $16\times16\times4$ obtains a higher Dice score of 1.38\% (p\textless0.005) than that of $16\times16\times8$, which indicates larger patch size may not always benefit the final performance.
Compared Fig. \ref{fig_tree_graph_seg_result_quantity}(c) and Fig. \ref{fig_tree_graph_seg_result_quantity}(d), we can find an interesting phenomenon that hat the input size has a very different effect on the performance. 
For tree data based segmentation, the difference between the performance of the implementation with the high-resolution and that with the low-resolution input is only 0.12\% (p\textgreater0.05).
While for graph based segmentation, the implementation with the high-resolution input outperforms that with the low-resolution inputs by 2.95\% (p\textless0.0001) in Dice score.

We also discuss the failed cases in pre-segmentation which is a critical step for both tree data based segmentation and graph based segmentation as shown in Fig. \ref{fig_tree_graph_seg_result_visual}.
Note that the two methods are both centerline-based solutions, relying on the centreline to construct data and segment the vessels.
In practical terms, we lack the labels of the true centreline to support the training of a centreline model.
Here, we adopt the segmentation of the network which are then labelled and skeletonised to make an approximate centreline to implement both methods.
Therefore, the quality of the pre-segmentation becomes a key factor in the accuracy of the subsequent segmentation.
As shown in Fig. \ref{fig_tree_graph_seg_result_visual},  some coronary arteries are missing in pre-segmentation, and as a result the missing vessels will not be constructed as a tree structure or graph structure.
Finally, these coronary arteries are missing in the rest of the processing, resulting in errors in the final segmentation.

\subsubsection{Baseline method}

We discussed several factors including modules (patch segmentation and coarse segmentation), ensemble, dilation (w/ and w/o), and patch size (16$^3$, 32$^3$, and 64$^3$).
The performance is shown in Table \ref{baseline_results}.
For the patch segmentation module, Dice scores of 79.56\%, 81.22\% and 82.34\% are obtained without dilation for patch sizes of 16$^3$, 32$^3$ and 64$^3$, respectively, and their paired comparisons (16$^3$ vs 32$^3$ (p\textless0.0001), 16$^3$ vs 64$^3$ (p\textless0.0001), and 32$^3$ vs 64$^3$ (p\textless0.001)) show statistical significance.
We can notice that larger patch sizes indicate larger receptive field, which benefits the segmentation.
While with dilation, larger patch sizes still obtains higher Dice score. 
However, there is no statistical significance (p\textgreater0.05) between a patch size of 32$^3$ and that of 64$^3$.
The main reason is that dilation is also effective to extract the context information.
The combination of the patch size of 32$^3$ and dilation is powerful enough to extract the context information, and larger receptive field with a larger patch size and dilation cannot further extract more context information.
We can also note that dilation can improve the performance with the patch size of 64$^3$ and 32$^3$ but not for 16$^3$, which may due to the fact that dilation will dismiss some pixels which is with a small quantity in the input with a small patch size. 
In terms of ensemble, the ensemble using the coarse segmentation module with the patch segmentation module (82.96\%) outperforms that of each base classifier (77.80\%, 82.27\%, 82.70\%) in terms of Dice score.

\begin{table}[t]\small
	\centering
	\caption{Dice score (in \%) of the baseline method. Intermediate results including those from the coarse segmentation step and the branches of the patch segmentation are also included.
	}
	\label{baseline_results}
	\begin{tabular}{p{3.2cm}p{1.1cm}p{1.4cm}p{1.4cm}}
		\toprule[2pt]
		\textbf{Module} &\textbf{Dilation} & \textbf{Patch} \textbf{Size} & \textbf{Dice} \textbf{score}  \\
		\midrule[2pt]
		 Patch segmentation &w/o  & 16$^3$ &  79.56  \\
		 Patch segmentation &w/o & 32$^3$ &  81.22\\
		 Patch segmentation &w/o & 64$^3$ &  82.34\\
		 Ensemble (Patch) &w/o & 16$^3$,32$^3$,64$^3$ & 80.97  \\
		 Ensemble (Coarse+Patch)& w/o & 16$^3$,32$^3$,64$^3$ & 82.21\\
		 Patch segmentation &w/  & 16$^3$ &  77.80 \\
		 Patch segmentation &w/ & 32$^3$ &  82.27 \\
		 Patch segmentation &w/ & 64$^3$ &  82.70  \\
		 Ensemble (Patch) &w/ & 16$^3$,32$^3$,64$^3$ & 81.11 \\
		 Ensemble (Coarse+patch) &w/ & 16$^3$,32$^3$,64$^3$ & 82.96 \\
		 		\bottomrule[2pt]
	\end{tabular}
\end{table}

\begin{figure*}[!htb]
	\centering
	\includegraphics[width=0.99\textwidth]{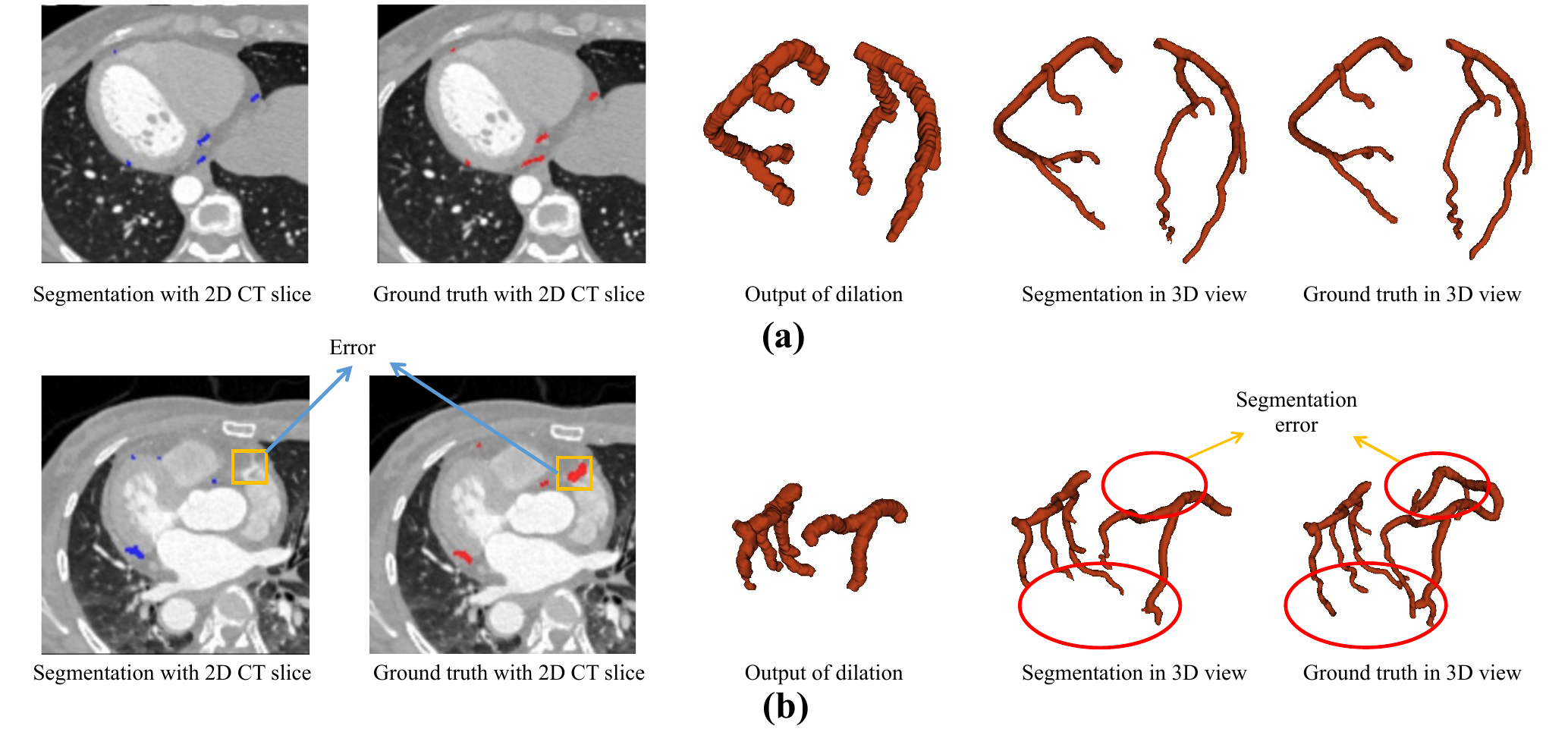}
	\caption{
        \textcolor{ForestGreen}{Visual discussion of (a) good and (b) poor segmentation cases in the baseline method.}
	} \label{fig_baseline_seg_result_visual}
\end{figure*}

Visual illustration of good and poor segmentation cases is shown in Fig. \ref{fig_baseline_seg_result_visual}.
As shown in Fig. \ref{fig_baseline_seg_result_visual}(a), the segmentation result is good which can well match the ground truth in both the 2D CT slices and the 3D view.
We can also notice that the output of dilation covers all the areas of the ground truth.
Fig. \ref{fig_baseline_seg_result_visual}(b) shows an example of poor segmentation.
We can notice that a thick vessel and a long thin vessel are missing which is due to the fact that they are not recognized by the dilated vessel segmentation module.
Particularly, the thick vessel as shown in the 2D CT slice is close to the right atrium and has a similar grey scale value to the right atrium.
As a result, the vessel is relatively hard to be recognized correctly. 


\begin{figure*}[!hbt]
\centerline{\includegraphics[width=1\textwidth]{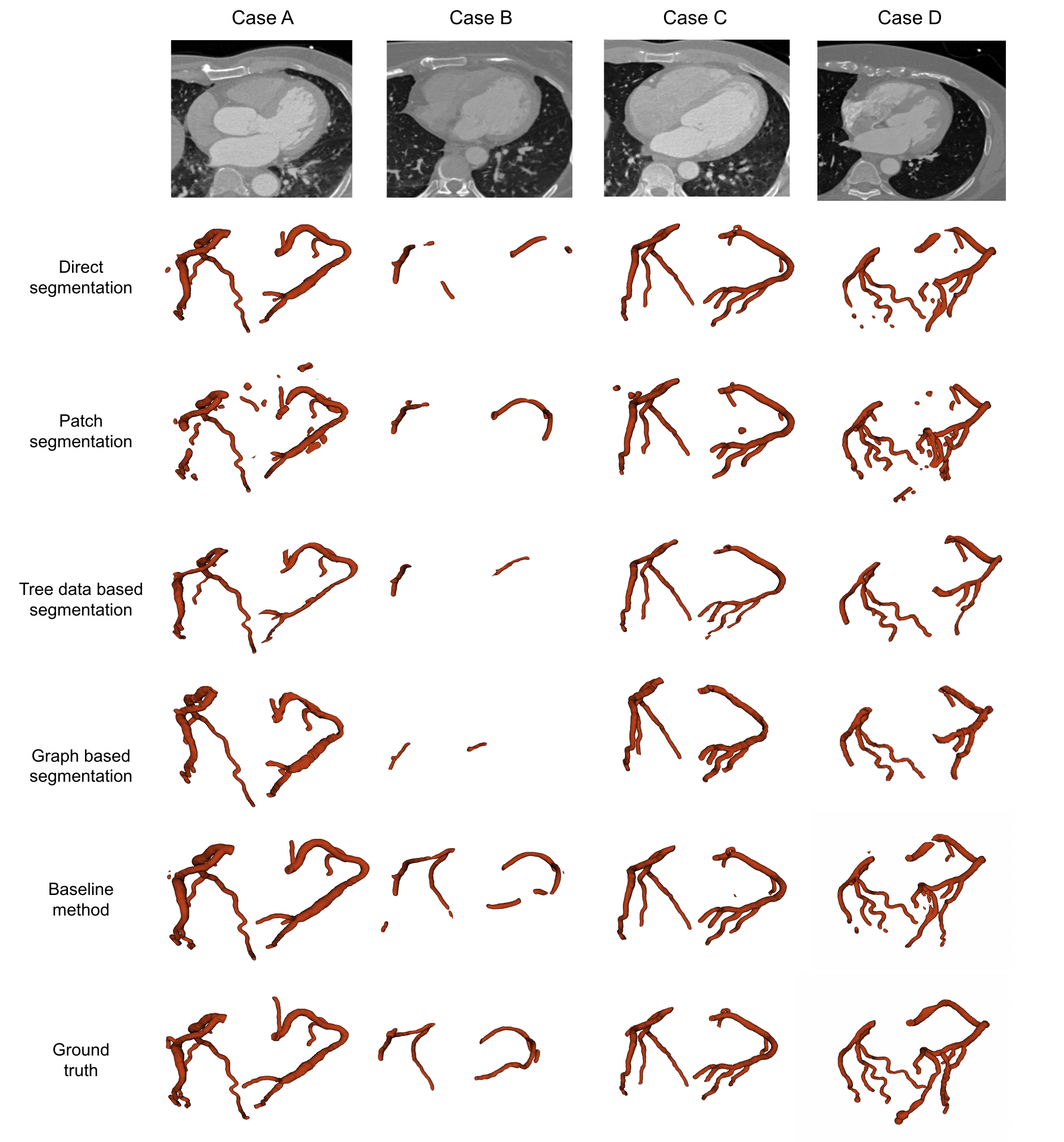}}
\caption{Visual comparison of the results of the five methods in the benchmark with four cases.}
\label{fig_all_seg_result_visual}
\end{figure*}

\subsection{Benchmark comparison}
\begin{table*}[t]\small
	\centering
	\caption{Performance comparison of the methods in the benchmark in Dice score (\%). Each method is with the optimal result which is selected from implementations with various configurations. 
	}\label{benchmark_results}
	\begin{tabular}{p{5cm}p{2cm}p{3cm}p{2cm}p{1.4cm}p{1.6cm}}
		\toprule[2pt]
		Method & Input type &Input size & Dice score (\%) & HD (mm) & AHD (mm)  \\
		\midrule[2pt]
		 Direct segmentation (3D FCN) \cite{shen2019coronary}  & Full Image & $512\times 512 \times 256 $ & 80.58 & 28.6656 & 0.8503 \\
		 Patch segmentation (3D U-net) \cite{huang2018coronary,chen2019coronary} & Patch & $64\times64\times64 $ & 72.01 & 40.9693 & 3.0686 \\
		 Tree data based segmentation (3D TreeConvGRU) \cite{kong2020learning} & Tree & $N\times16\times16\times4 $ &  68.78 & 30.3350 & 1.4336 \\
		 Graph based segmentation (GCN) \cite{wolterink2019graph} & Graph & $N\times32 $ & 70.61 & 27.8718  & 1.2433\\
		 Baseline method (3D U-net and U-net++) & Patch & $128\times128\times64$, 16$^3$,32$^3$,64$^3$ & 82.96 & 27.2169 & 0.8180\\
		 		\bottomrule[2pt]
	\end{tabular}
\end{table*}

Performance comparison of all the methods in the benchmark is shown in Table \ref{benchmark_results}. 
We can notice that the proposed baseline method achieves the optimal performance on all metrics.
We can also discover an interesting phenomenon that patch segmentation, tree data based segmentation and graph based segmentation has a much lower performance than direct segmentation.
We can make a rough discussion here.
Such phenomenon partially shows the same trend in existing works in which direct segmentation (89 in \cite{cheung2021computationally}) has higher Dice score than patch segmentation (94 in \cite{pan2021coronary}, tree data based segmentation (85 in \cite{kong2020learning}), and Graph based segmentation (73-75 in \cite{wolterink2019graph}).
But we can still notice some inconsistencies that patch segmentation has a higher performance than direct segmentation in existing works.
There are mainly two reasons.
First, the dataset for evaluation is not the same, and the quanlity of the datasets also varies.
Note that our dataset is the largest one currently, which is orders of magnitude larger than most of the datasets used by existing methods.
Second, there are much details including hyper parameters, pre-processing and post-processing in the existing works which is relatively hard for implementation.
Though we have tried our best to implement existing works, there is some details (sometimes maybe critical details) that are inevitably missed due to the lack of understanding and the limited length of the related papers.
Thus, we welcome related researchers in the community to join us to improve the implementations.

Visual comparison of the methods in the proposed benchmark is shown in Fig. \ref{fig_all_seg_result_visual}.
For direct segmentation, the performance of Case A and Case C is good.
However, when the quality of the images is low and the structure of the coronary arteries varies considerably as shown in Case B and D, areas with low contrast cannot be correctly detected.
For patch segmentation, the results are quite similar to that of direct segmentation.
The only difference is that patch segmentation can recognize some thin vessels in Case D but cannot in Case A, while direct segmentation has a opposite performance on the vessels in the two cases.
Also patch segmentation has a slightly better performance than that of direct segmentation in Case B.
It seems that patch segmentation can process local features better than direct segmentation patch which is expected as patch segmentation pay more attention on local feature processing.
Tree data based segmentation and graph based segmentation share a rather similar, however, they cannot discover the vessels with low contrast well in Case B and Case D.
The main reason is that tree data based and graph based segmentation methods rely heavily on the centreline extracted using the pre-segmentation, which determines the number of nodes in the tree and graph structure.
For the proposed baseline method, the results show much better performance especially on Case B where the majority of the vessels are correctly recognized.
The main reason is that the baseline method combines both the features from coarse segmentation and patch segmentation with various patch sizes, which can better extract the context information.

\subsection{\textcolor{ForestGreen}{Discussion}}
\textcolor{ForestGreen}{
We have implemented several typical deep learning based methods including direct segmentation \cite{shen2019coronary}, patch based segmentation \cite{huang2018coronary,chen2019coronary},
tree data based segmentation \cite{kong2020learning}, graph data based segmentation \cite{wolterink2019graph}, and our baseline method.
The proposed baseline method achieves the optimal performance on Dice score, HD and AHD. 
On the other hand, we can also notice that the tree data based segmentation method achieves a better performance than direct segmentation, which is contrary to the results in \cite{kong2020learning}.
This may be partially due to the fact that there are many technique details in the training and optimization in \cite{kong2020learning}.
This is also the reason that we published our dataset and code for fair comparison.
}

\textcolor{ForestGreen}{Though our dataset is quite large compared with exiting works and well labeled by two to three experienced radiologists, it has limitations.
First, our dataset was collected in only one center thus with biased distributions.
Second, only one kind of CT machines namely Siemens 128-slice dual-source scanner was used to acquire the CT images, which make the biased problem even worse.
Third, no detailed labels is provided. For example, subclassess of coronary artery including left main coronary artery, left anterior descending coronary
artery, etc. are not separated.
We hope that others can also publish their datasets to mitigate the above limitations and at the same time facilitate related research.
}

\textcolor{ForestGreen}{Future directions of our benchmark can be various, and we only name a few here.
First, more advanced segmentation networks like nnU-net \cite{isensee2021nnu}, CoTr \cite{xie2021cotr}, and UNETR \cite{hatamizadeh2022unetr} can be employed in existing frameworks to enhance performance.
Second, more advanced networks and evaluation metrics can be explored to preserve the topology or connectivity \cite{shit2021cldice, hu2019topology,hu2021topology,saeki2021statistical} of the coronary vessels which plays a critical role for further analysis and diagnosis.
Third, based on our dataset and existing ones \cite{schaap2009standardized,kiricsli2013standardized}, multi-center related topics including federal learning \cite{rajasree2022role} and domain adaptation \cite{guan2021domain} can be further investigated.
}

\section{Conclusion}

\textcolor{black}{Segmentation of coronary arteries is a critical task for the diagnosis and quantification of coronary artery disease. 
In this paper, we propose a benchmark dataset for coronary artery segmentation on CTA images. In addition, we have implemented a benchmark in which we have not only tried our best to implement several typical existing methods but also proposed a strong baseline method. 
We have performed a comprehensive evaluation of the methods in the benchmark, and the results show that the proposed baseline method achieves the optimal performance with a Dice score of 82.96\%.
However, the performance still has room of improvement for accurate diagnosis and stenosis quantification for clinical practice.
The benchmark and the dataset is published at https://github.com/XiaoweiXu/ImageCAS-A-Large-Scale-Dataset-and-Benchmark-for-Coronary-Artery-Segmentation-based-on-CT. 
We hope that the proposed dataset and benchmark can stimulate further research in the community.
}

\textcolor{ForestGreen}{\section{Ethical and information governance approvals}
This work and the collection of data of retrospective data on implied consent received Research Ethics Committee (REC) approval from Guangdong Provincial People’s Hospital, Guangdong Academy of Medical Sciences under Protocol No. 2019324H. It complies with all relevant ethical regulations. Deindentification was performed in which all CT files are transformed into NIfTI format, and sensitive information of the patients including name, birther day, admission year, admission number, and CT number is removed. 
}

\section{Acknowledgments}
\textcolor{black}{
This work was supported by the Science and Technology Planning Project of Guangdong Province, China (No. 2019B020230003), Guangdong Peak Project (No. DFJH201802), the National Natural Science Foundation of China (No. 62006050, No. 62276071), Science and Technology Projects in Guangzhou, China (No. 202206010049, No. 2019A050510041), Guangdong Basic and Applied Basic Research Foundation (No. 2022A1515010157, 2022A1515011650), and Guangzhou Science, Technology Planning Project (No. 202102080188), and the Sanming Project of Medicine in Shenzhen (No. SZSM202011005).
}

\bibliographystyle{abbrv}
\bibliography{reference}


\end{document}